\documentclass[aps,pre,twocolumn,groupedaddress,showpacs]{revtex4-2}
\newcommand{\bec}[1]{\mbox{\boldmath $ #1$}}
\usepackage{bm}
\usepackage{graphicx}
\usepackage{color}

{}
\newcommand{\meanUU}{\overline{\bm{U}}}

{}
{}

\newcommand{\meanN}{\overline{n}}

\newcommand{\meanT}{\overline{T}}

\newcommand{\meanU}{\overline{U}}

\begin{document}
\title{Experimental investigation of turbulence and turbulent thermal diffusion
in strongly inhomogeneous and anisotropic forced convection}
\author{E.~Zarbib}
\author{E.~Elmakies}
\author{O.~Shildkrot}
\author{N.~Kleeorin}
\author{A.~Levy}
\author{I.~Rogachevskii}
\email{gary@bgu.ac.il}

\bigskip
\affiliation{
The Pearlstone Center for Aeronautical Engineering
Studies, Department of Mechanical Engineering,
Ben-Gurion University of the Negev, P.O.Box 653,
Beer-Sheva 8410530,  Israel}

\date{\today}
\begin{abstract}
We investigate properties of turbulence and
turbulent transport of non-inertial particles described in terms of
turbulent thermal diffusion in strongly inhomogeneous and anisotropic convection
forced by two similar turbulence generators with oscillating membrane and a steady grid in the air flow
(with the Rayleigh number about $10^8$).
Velocity field and spatial distribution of
particles are measured using Particle Image  Velocimetry system.
The temperature distribution is measured in many locations
using a temperature probe equipped with 12 E - thermocouples.
In the forced convection, the gradients of the mean temperature field and the particle
number density in the horizontal direction in the core flow are much
stronger than in the vertical direction.
The mean fluid velocity structure show transition between a single-roll pattern for isothermal
turbulence to double-roll patterns with increase of the temperature difference between the bottom
and upper walls of the chamber.
For larger temperature differences, the mean fluid velocity structure returns to
a single-roll pattern.
In the turbulent regions with large mean temperature gradients, the dominant effect
of the large-scale particle clustering is turbulent thermal diffusion, resulting in that
the maximum of the mean particle number density is located in the regions with minimum
of the mean temperature and vise versa.
Deviations from this feature is observed in the regions with strong mean fluid velocities
where the mean temperature gradients are small.
\end{abstract}

\maketitle

\section{Introduction}
\label{sect1}

Turbulent transport of particles has been studied in a number of papers due to a large variety of applications
in geophysical,  astrophysical and industrial flows
\cite{CSA80,ZRS90, BLA97, SP06, ZA08,CST11,RI21}.
Important effects in turbulent transport of particles are formation of small-scale and large-scale clusters
in particle spatial distributions.
In a fully developed turbulence, characteristic sizes of small-scale clusters are much smaller than the integral turbulence scale $\ell_0$,
while large-scale clusters have sizes larger than $\ell_0$.
Similarly, characteristic life-time of small-scale clusters are much smaller than the turbulent time $\tau_0$ in the integral turbulence scale,
whereas  life-time of large-scale clusters are of the order of turbulent diffusion time which is much larger than $\tau_0$
 \cite{SP06,ZA08,CST11,RI21,WA00,S03,G08,WA09}.

 In an isothermal inhomogeneous turbulence, one of the important mechanism of large-scale particle clusters
 is turbophoresis that is combined effect of particle inertia and inhomogeneity of turbulence.
 This phenomenon causes the formation of large-scale particle clusters in the vicinity of the
 minimum turbulence intensity \cite{CTT75,RE83,G97,EKR98,G08,SSB12,LCB16,MHR18}.

 In density or temperature stratified turbulence,
 large-scale particle clusters are formed due to turbulent thermal diffusion
 \cite{EKR96,EKR97,RI21} that
 occurs for non-inertial and inertial particles.
The latter effect causes a non-zero effective pumping
velocity in the turbulent flux particles in the direction opposite
to the mean temperature gradient, so that particles are accumulated in the vicinity of the mean
temperature minimum.

Below we shortly discuss the mechanism of  turbulent thermal diffusion
resulting in the large-scale clustering of small non-inertial particles
in a developed turbulence.
Equation for the particle number density $n(t,{\bm x})$ advected by
fluid velocity field ${\bm U}(t,{\bm x})$ is given by
\begin{eqnarray}
{\partial n \over \partial t} + {\bm \nabla} {\bf \cdot}({\bm U}\, n) = D \, \Delta n ,
\label{W1}
\end{eqnarray}
where $D= k_B \,T/(3\pi \rho \, \nu \, d_{\rm p})$ is the coefficient
of the molecular (Brownian) diffusion of particles,   $d_{\rm p}$ is the particle diameter,
$\nu$  is the kinematic viscosity of the fluid, $T$ and $\rho$  are the fluid temperature and
density and $k_B$ is the Boltzmann constant.
We consider a low-Mach-number fluid flow with inhomogeneous density, so that
the continuity equation in an anelastic approximation is given by
$\bec{\nabla} {\bf \cdot}(\rho \, {\bm U}) = 0$.

To describe the large-scale clustering of particles,
we apply a mean-field approach and use Reynolds averaging,
so that all quantities are decomposed into the
mean and fluctuating parts, and the fluctuating parts have zero mean values.
Averaging Eq.~(\ref{W1}) over an ensemble of turbulent velocity field,
we arrive at the mean-field equation for the particle number density:
\begin{eqnarray}
{\partial \overline{n} \over \partial t} + {\bm \nabla} {\bf \cdot} \Big(\meanUU \, \meanN +\langle {\bm u} \, n'  \rangle \Big) = D \, \Delta \overline{n} ,
\label{W2}
\end{eqnarray}
where $\overline{n}=\langle n \rangle$ is the mean particle number density, $n'$ are particle number density fluctuations,
$\langle {\bm u} \, n' \rangle$ is the turbulent flux of particles,
${\bm u}$ are velocity fluctuations, and the angular brackets denote ensemble averaging.
The particle turbulent flux $\langle  {\bm u} \, n'  \rangle$ is given by  \cite{EKR96,EKR97}
\begin{eqnarray}
\left\langle {\bm u} \, n'  \right\rangle = {\bm V}^{\rm eff} \, \overline{n} - D_{\rm T} \, {\bm \nabla} \overline{n} ,
\label{W7}
\end{eqnarray}
where $D_{\rm T}$ is the turbulent diffusion coefficient and ${\bm V}^{\rm eff}$ is
the effective pumping velocity that for small non-inertial particles is given by
\begin{eqnarray}
{\bm V}^{\rm eff}  = D_{\rm T} \, {{\bm \nabla} \overline{\rho} \over \overline{\rho}} \approx
- D_{\rm T} \, {{\bm \nabla} \meanT \over \meanT} .
\label{W8}
\end{eqnarray}
Here $\overline{\rho}$ and $\meanT$ are the mean fluid density
and temperature, respectively, and
we take into account that for a small mean pressure gradient,
${\bm \nabla} \, \ln\overline{\rho} \approx  - {\bm \nabla} \ln \meanT$.
Note that various rigorous methods yield Eqs.~(\ref{W7})--(\ref{W8}),
which implies that this result is robust (see, e.g., Ref.~\cite{RI21}).
For illustration, the equation for the turbulent flux of particles $\langle {\bm u} \, n' \rangle$
for an isotropic turbulence is derived in Appendix.
Using Eqs.~(\ref{W7})--(\ref{W8}), we rewrite Eq.~(\ref{W2}) for the mean particle number density as
\begin{eqnarray}
{\partial \meanN \over \partial t} + {\bm \nabla} {\bf \cdot} \biggl[\biggl(\meanUU - D_{\rm T} \, {{\bm \nabla} \meanT \over \meanT}\biggr) \meanN - D_{\rm T} {\bm \nabla} \meanN\biggr]  = 0 ,
\label{W10}
\end{eqnarray}
where we neglect small molecular (Brownian) diffusion coefficient $D$ for large Peclet numbers.
Note that Eqs.~(\ref{W7})--(\ref{W8}) are also valid when the mean velocity does not vanish.
We assume that the gradients along the horizontal axis $Y$ of the mean temperature, mean velocity,
mean number density and turbulent diffusion coefficient are much larger than those in other directions.
This condition corresponds to that observed in the experiments discussed in this paper.
Thus, the steady-state solution for Eq.~(\ref{W10}) for a zero total particle flux at the boundaries
reads
\begin{eqnarray}
{\meanN(Y) \over \meanN_0} =  \biggl({\meanT(Y) \over \meanT_0}  \biggr)^{-1} \, \exp \biggl(\int_0^Y
{\meanU_y \over D_{\rm T}} \,{\rm d} Y' \biggr) ,
\label{W11}
\end{eqnarray}
where $\meanN(Y=0) = \meanN_0$ and $\meanT(Y=0) = \meanT_0$.
Equation~(\ref{W11}) implies that maximum particle number density is attained near the region
with minimum mean fluid temperature.

\begin{figure*}[t!]
\centering
\includegraphics[width=8.0cm]{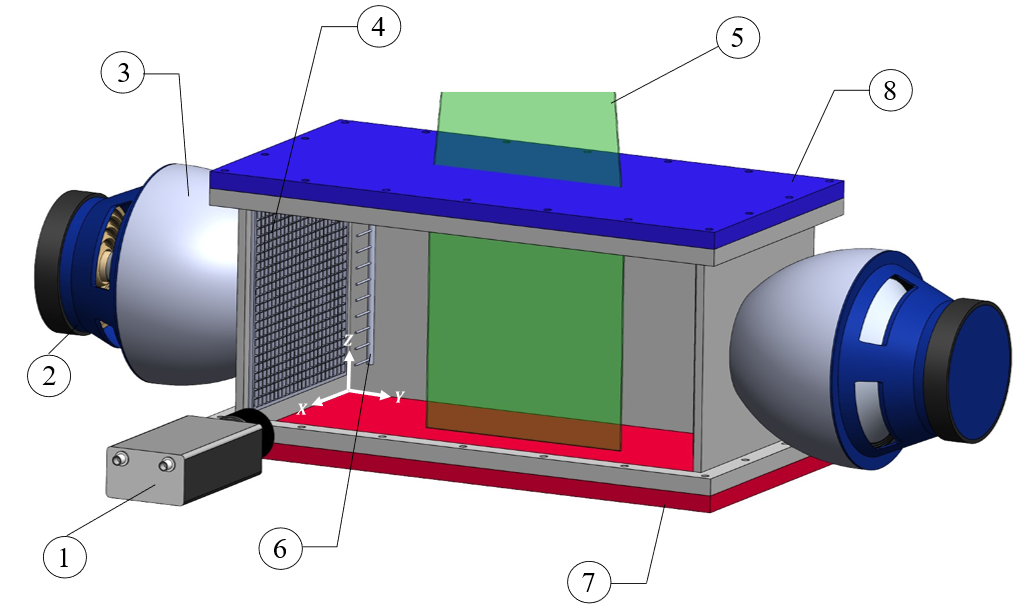}
\includegraphics[width=8.0cm]{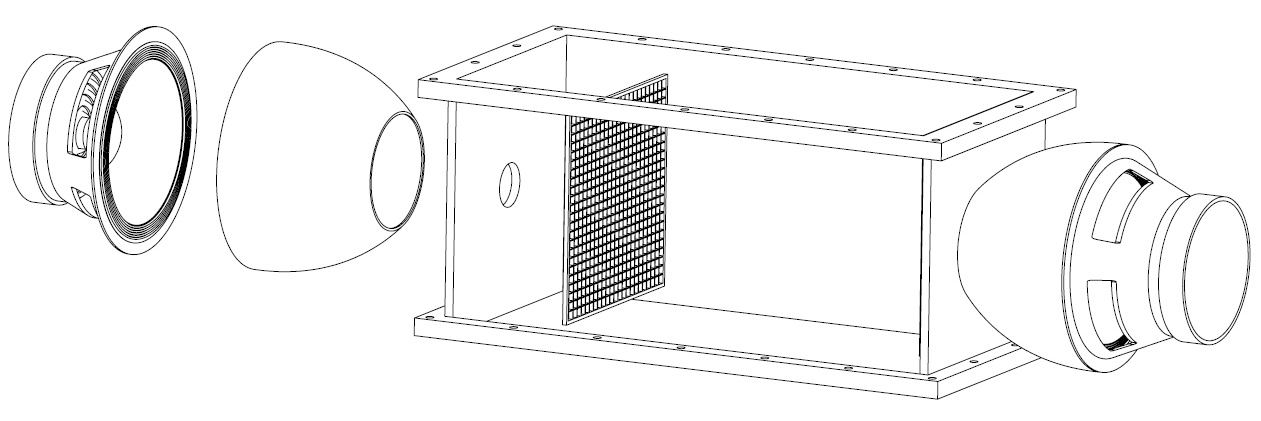}
\caption{\label{Fig1}
Experimental setup  with the forced convective turbulence (left panel)
produced by two turbulence generator with oscillating membrane and a steady grid:
(1) digital CCD camera;
(2) and (3) turbulence generator with oscillating membrane;
(4) steady grid;
(5) laser light sheet;
(6) temperature probe equipped with 12 E - thermocouples;
(7) heat exchanger at the bottom heated wall of the chamber;
(8) heat exchanger at the top  cooled wall of the chamber.
Semi-exploding view of the experimental system, showing the different components
assembly  (right panel).
}
\end{figure*}

Phenomenon of turbulent thermal diffusion has been studied analytically
applying various theoretical methods (e.g., the dimensional analysis,
the $\tau$ approaches applied in physical and Fourier spaces,
the path integral approach, the direct interaction approximation and functional approach, see
\cite{RI21,EKR96,EKR97,EKR00,EKRS00,EKRS01,PM02,RE05,AEKR17}).
This effect has been
detected in direct numerical simulations (DNS) \cite{HKRB12,RKB18},
observed in geophysics \cite{SSEKR09},
and discussed in planetary \cite{EKPR97} and astrophysical turbulence applications \cite{H16,KR25}.

Turbulent thermal diffusion has been detected
in laboratory experiments with micron-size particles
in temperature stratified turbulence
produced by one or two oscillating grids
\cite{BEE04,EEKR04,EEKR06a,AEKR17,EKRL22,EKRL23},
or by multi-fan turbulence generator \cite{EEKR06b}.
This phenomenon was even observed for nanoparticles \cite{SKRL22}
in temperature stratified turbulence.
All these experiments show the formation of large-scale particle clusters
in the regions with the mean temperature minimum
caused by turbulent thermal diffusion.

In the present paper we study in laboratory experiments
properties of turbulence and a large-scale clustering of small solid particles
in a convective turbulence forced by two similar turbulence generators
which contain both, oscillating membrane and a steady grid in the air flow.
This forcing is completely different from that applied to turbulent convection
in the previous experiments \cite{EEKR04,EEKR06a,EKRL23}, where the large-scale circulations
in turbulent convection were destroyed by the applied forcing, so that
the mean velocity fields were much smaller than velocity fluctuations.
In the present convective experiments, the forcing produces very complicated large-scale velocity field
with large-scale circulations and strong horizontal jets.
As the result, the gradients of the mean temperature and the mean
particle number density in the horizontal direction occur to be much
larger than in the vertical direction.
Thus, in the first time,  the large-scale clustering of small solid particles
is investigated in the horizontal direction rather than in the vertical direction.

This paper is organized as follows.
In Sec. ~\ref{sect2} we describe experimental setup
and measurement techniques.
In Sec. ~\ref{sect3} we analyse the obtained experimental results, and
in Sec.~\ref{sect4} we outline conclusions.
In Appendix we derive the equation for the turbulent flux of particles
for an isotropic turbulence.

\section{Experimental setup}
\label{sect2}

In this section we discuss the experimental set-up
and measurement technique.
We study properties of turbulence and a large-scale clustering of small solid particles
described in terms of the phenomenon of turbulent
thermal diffusion  in experiments with
a convective turbulence forced by two similar turbulence generators
with oscillating membrane and a steady grid  in the air flow.

The experiments are conducted in rectangular chamber with the sizes
$L_x \times L_y \times L_z$, where $L_x=L_z=26$ cm, $L_y=53$ cm.
Here the axis $Z$ is in the vertical direction and the axis $Y$ is perpendicular
to the oscillating membrane plain (see Fig.~\ref{Fig1}).
The membrane is driven using a LABVIEW based program designed for controlling
the frequency of the membrane motion (varying from 10 Hz to 2000 Hz) and the r.m.s voltage
(which affects the amplitude of the membrane motion).
In the experiments we mainly use the frequency of the membrane motion about 17 Hz.
The air converging galvanized steel ejector
(with 30 cm maximum diameter, 5 cm minimum diameter and 10 cm long)
is attached to the chamber wall, so the turbulent flow produced by the oscillating membrane
passes through the steady grid into the chamber.

A polypropylene cone woofer pushes air through  the ejector which increases
the mass flow rate and velocity of the air.
The outlet of the ejector is covered by a double layered Polymerized
Lactic Acid square adjustable grid (24 $\times$ 24 cm$^2$) with square array of 2-5 mm thick bars,
creating a mesh structure with the mesh size $9 \times 9$ mm$^2$,
which corresponds to the porosity constant $\approx 0.74$  (see Fig.~\ref{Fig1}).
For instance, at the frequency of 17 Hz of the oscillating membrane and r.m.s voltage of 2\, V,
and with a 9 mm mesh size grid, a measured turbulent velocity of up to 20 cm/s
is reached at 20 mesh size distance from the steady grid.

A vertical mean temperature gradient in the
turbulent flow is formed by attaching two
aluminium heat exchangers to the bottom  (heated) and top (cooled)
walls of the chamber which allow us to
form a mean temperature gradient in a turbulent flow.
The temperature field is measured in many locations applying a temperature probe
equipped with 12 E - thermocouples.
The thermocouples with the diameter of 0.13 mm and
the sensitivity of $\approx 75 \, \mu$V/K are attached to
a vertical rod with a diameter 4 mm, and the mean distance between thermocouples
is about 20 mm (see for details Ref.~\cite{EKRL22,EKRL23}).
The data are recorded using the developed software based
on LabView 2024 Q1, and the temperature maps are obtained
using Matlab~R2024a.

The velocity field is measured using a
Particle Image  Velocimetry (PIV) system \cite{AD91,RWK07,W00},
consisting in a Nd-YAG laser (Continuum Surelite $2 \times
170$ mJ) and a progressive-scan 12 bit digital CCD
camera (with pixel size $7.4 \, \mu$m $\times \,
7.4 \, \mu$m and $2048 \times 2048$ pixels).
As a tracer for the PIV measurements,
an incense smoke is used with spherical solid particles
having the mean diameter of $0.7 \mu$m
and the material density $\rho_{\rm p}\approx 10^3 \rho$.
The particles are produced by high temperature sublimation of solid
incense grains (see for details Ref.~\cite{EKRL22,EKRL23}).

\begin{figure*}[t!]
\centering
\includegraphics[width=17.0cm]{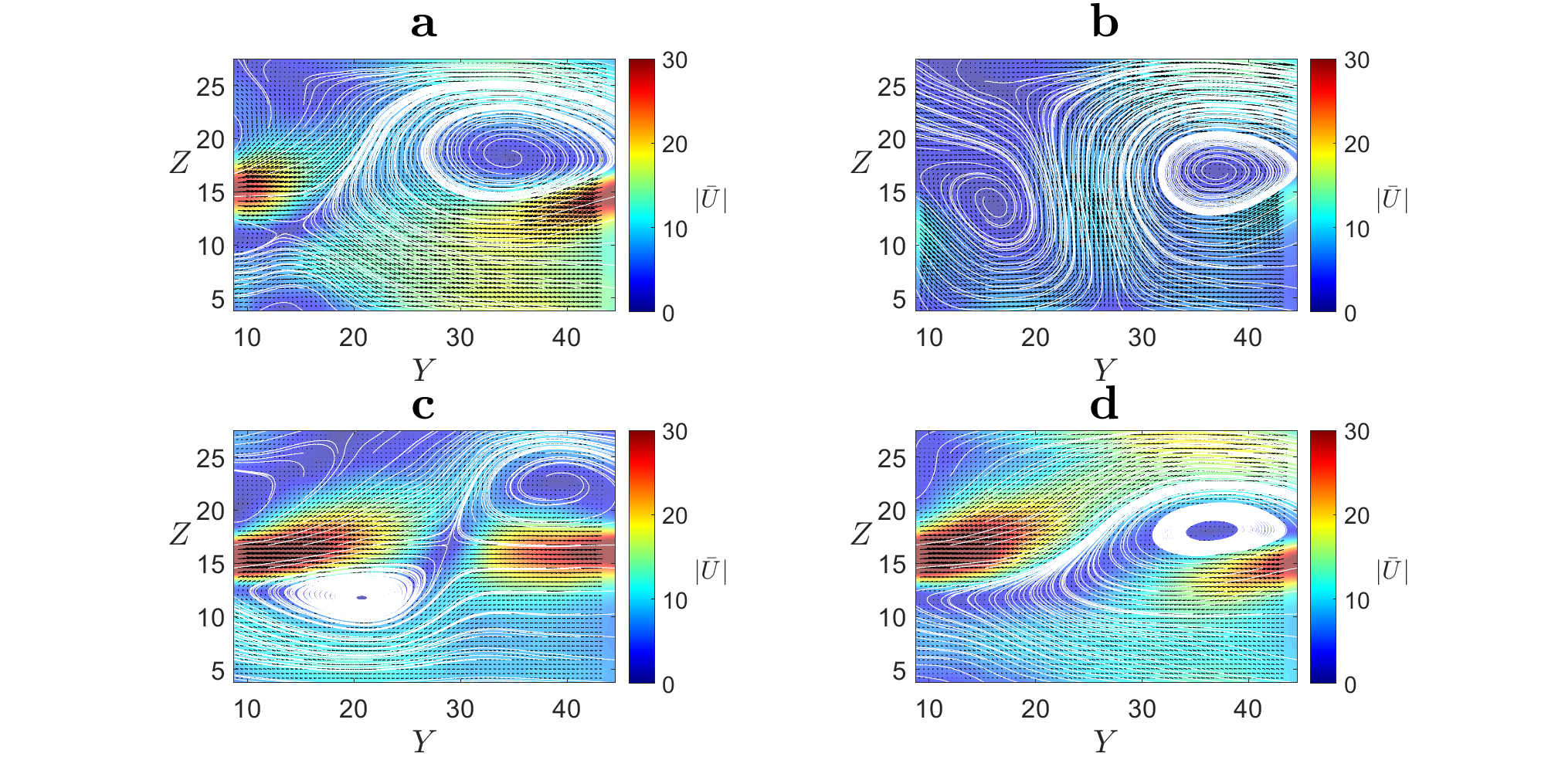}
\caption{\label{Fig2}
Distributions of the mean velocity field $|\meanU|$
in the core flow: {\bf (a)} for isothermal turbulence; and
for forced convective turbulence at the temperature differences:
{\bf (b)} $\Delta T = 40$~K; {\bf (c)} $\Delta T = 50$~K; and
{\bf (d)} $\Delta T = 60$~K between the bottom and upper walls of the chamber.
The velocity is measured in cm/s and coordinates $Y$ and $Z$ are measured in cm.
}
\end{figure*}

\begin{figure*}[t!]
\centering
\includegraphics[width=17.0cm]{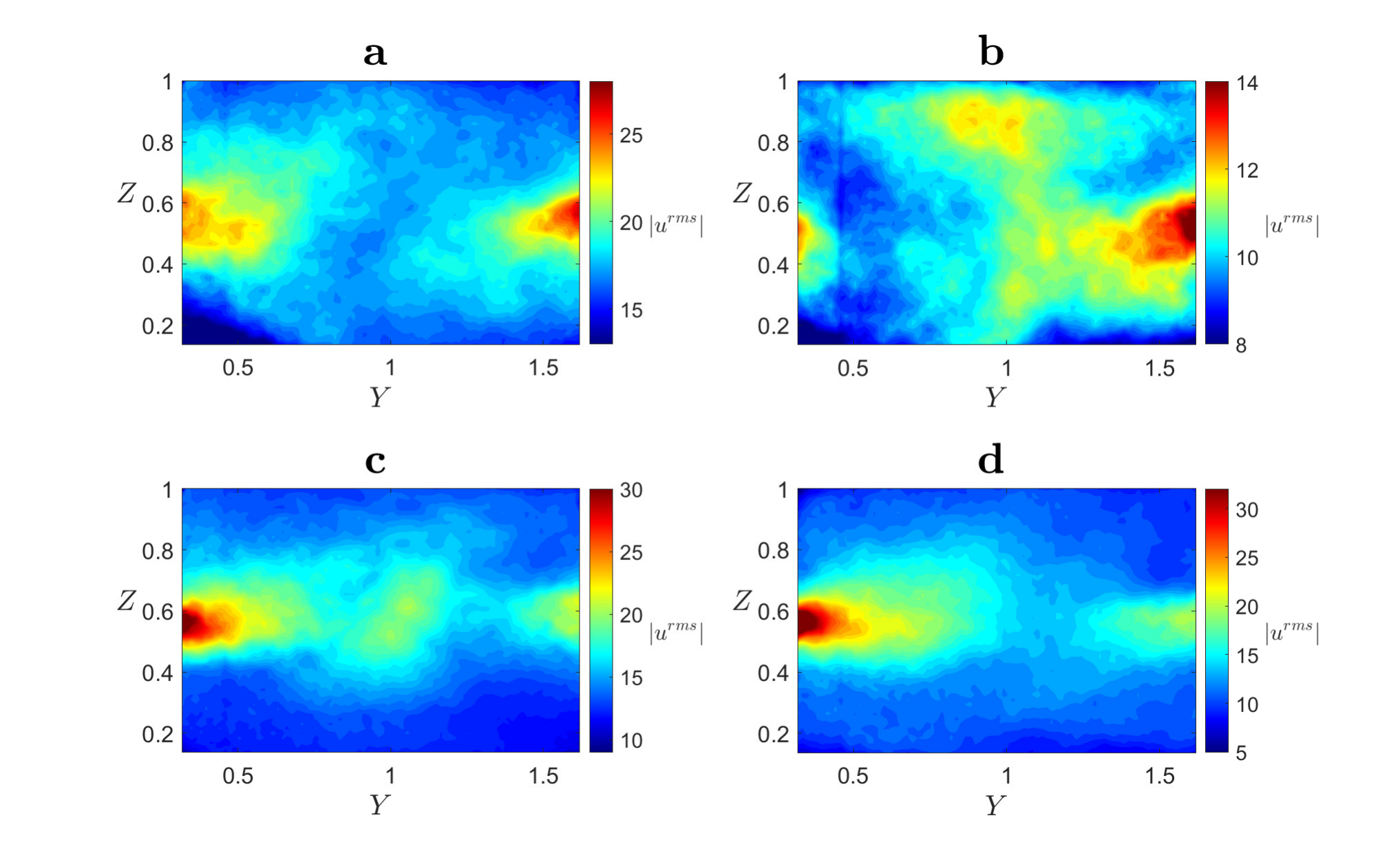}
\caption{\label{Fig3}
Distributions of the turbulent velocity $|u^{\rm (rms)}| = [\langle u_y^2 \rangle + \langle u_z^2 \rangle]^{1/2}$
in the core flow: {\bf (a)} for isothermal turbulence; and
for forced convective turbulence at the temperature differences:
{\bf (b)} $\Delta T = 40$~K between the bottom and upper walls of the chamber;
{\bf (c)} $\Delta T = 50$~K; and {\bf (d)} $\Delta T = 60$~K.
The velocity is measured in cm/s and coordinates $Y$ and $Z$ are normalized by $L_z=26$ cm.
}
\end{figure*}

\begin{figure*}[t!]
\centering
\includegraphics[width=17.0cm]{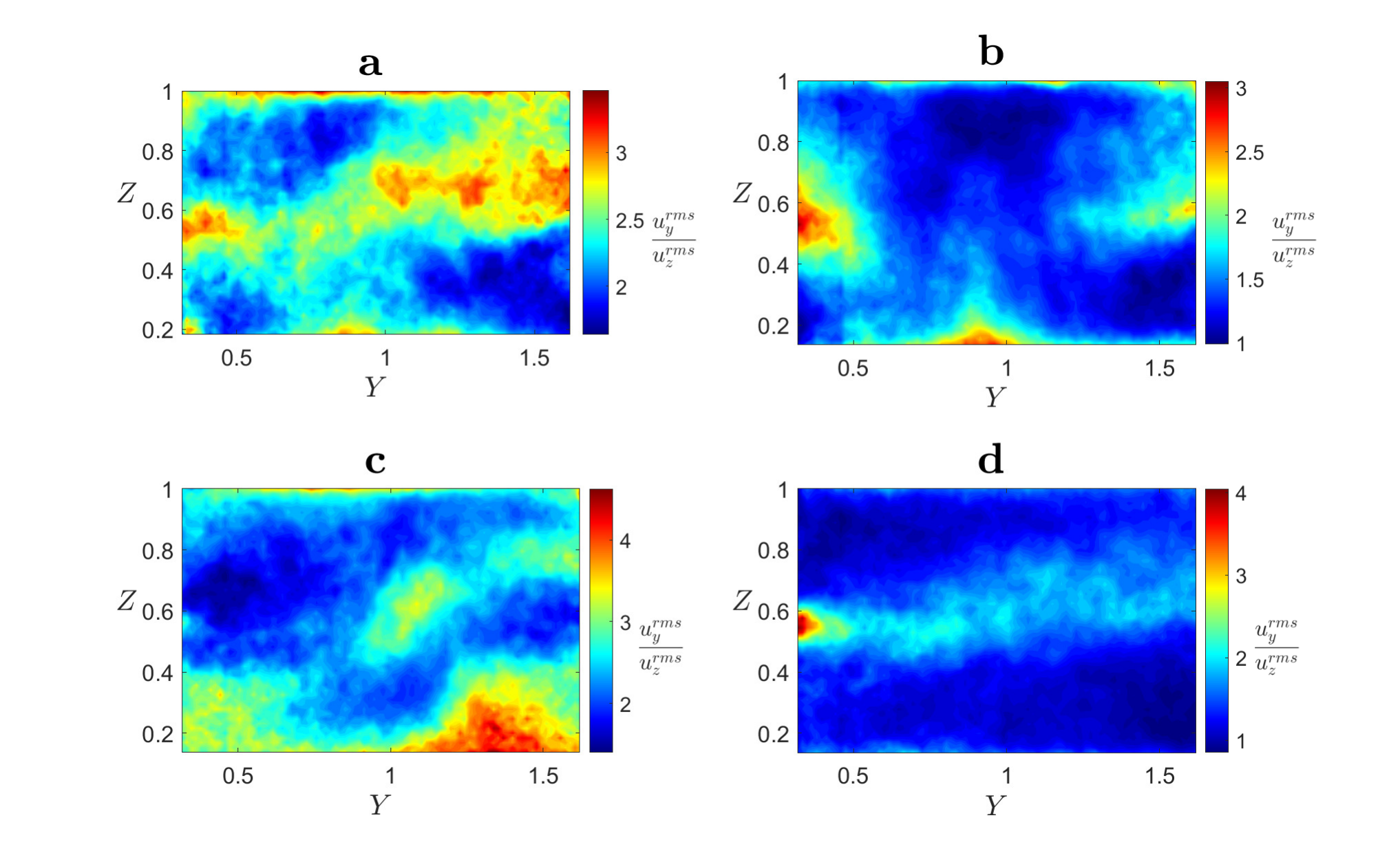}
\caption{\label{Fig4}
Distributions of the turbulent anisotropy $[\langle u_y^2 \rangle / \langle u_z^2 \rangle]^{1/2}$
in the core flow: {\bf (a)} for isothermal turbulence; and
for forced convective turbulence at the temperature differences:
{\bf (b)} $\Delta T = 40$~K between the bottom and upper walls of the chamber;
{\bf (c)} $\Delta T = 50$~K; and {\bf (d)} $\Delta T = 60$~K.
The velocity is measured in cm/s and coordinates $Y$ and $Z$ are normalized by $L_z=26$ cm.
}
\end{figure*}

\begin{figure*}[t!]
\centering
\includegraphics[width=17.0cm]{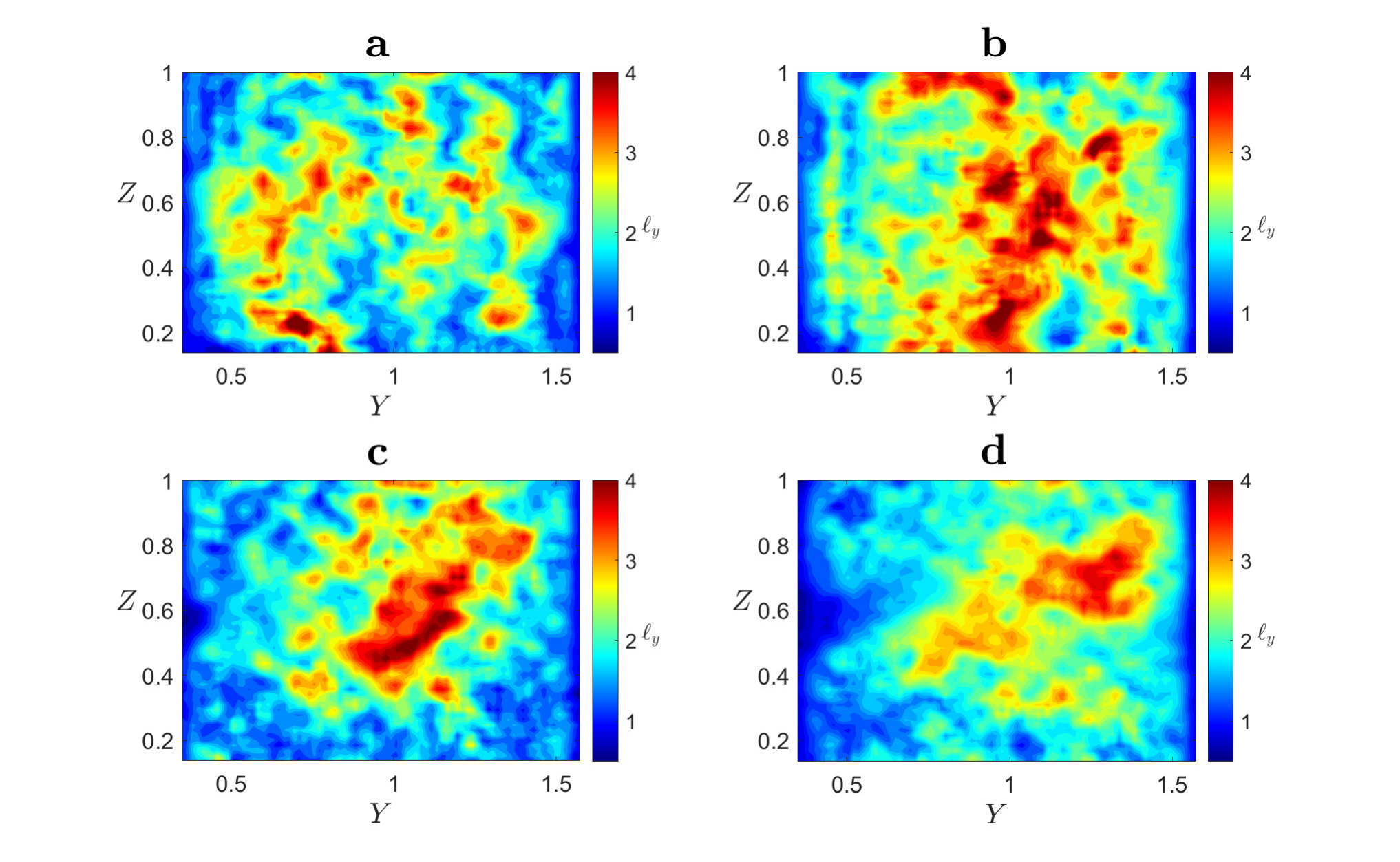}
\caption{\label{Fig5}
Distributions of  the horizontal  integral turbulent length scale $\ell_y$
in the core flow: {\bf (a)} for isothermal turbulence; and
for forced convective turbulence at the temperature differences:
{\bf (b)} $\Delta T = 40$~K between the bottom and upper walls of the chamber;
{\bf (c)} $\Delta T = 50$~K; and {\bf (d)} $\Delta T = 60$~K.
The integral turbulent length scale is measured in cm, and coordinates $Y$ and $Z$ are normalized by $L_z=26$ cm.
}
\end{figure*}

\begin{figure*}[t!]
\centering
\includegraphics[width=17.0cm]{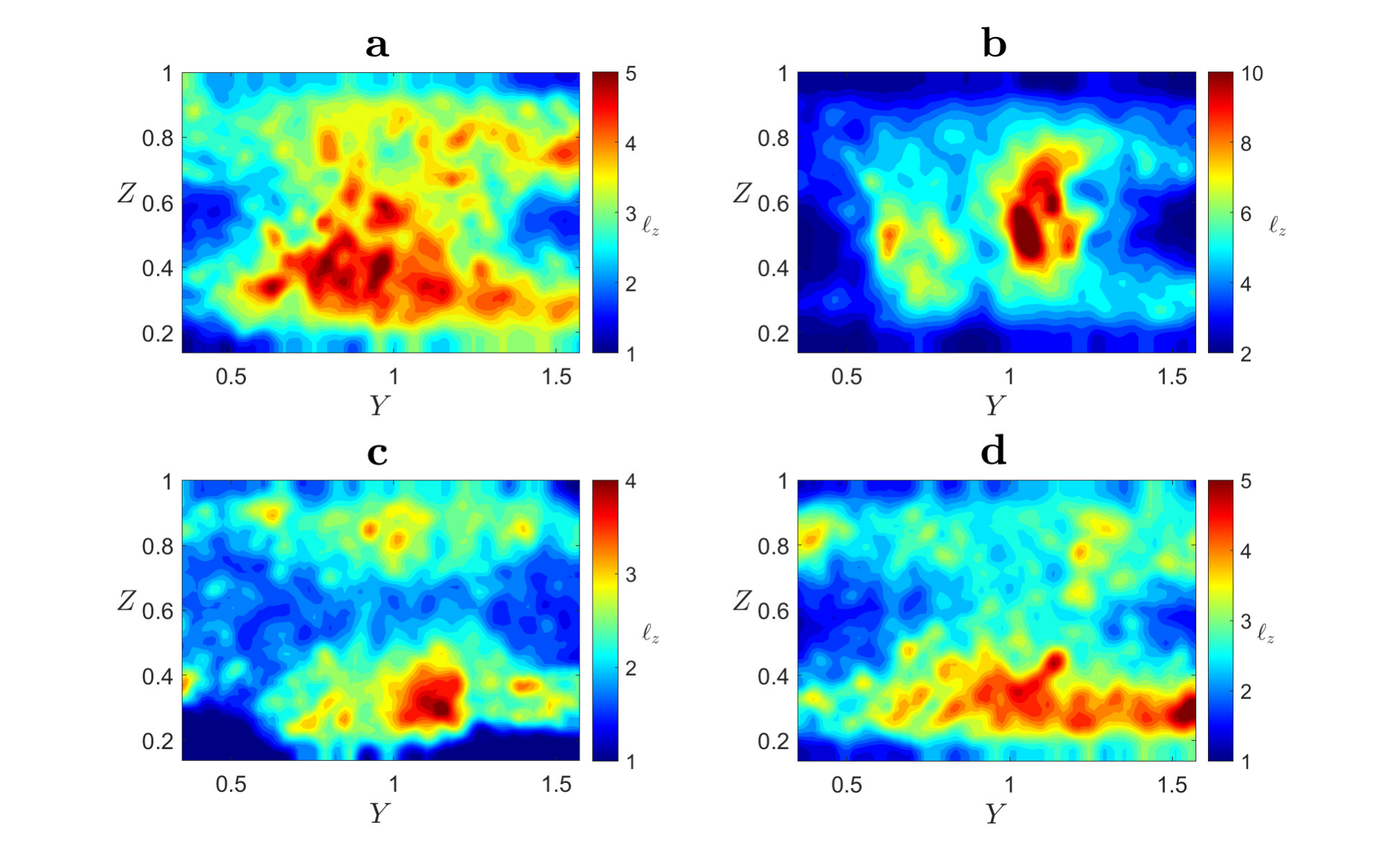}
\caption{\label{Fig6}
Distributions of  the vertical  integral turbulent length scale $\ell_z$
in the core flow: {\bf (a)} for isothermal turbulence; and
for forced convective turbulence at the temperature differences:
{\bf (b)} $\Delta T = 40$~K between the bottom and upper walls of the chamber;
{\bf (c)} $\Delta T = 50$~K; and {\bf (d)} $\Delta T = 60$~K.
The integral turbulent length scale is measured in cm, and coordinates $Y$ and $Z$ are normalized by $L_z=26$ cm.
}
\end{figure*}

\begin{figure*}[t!]
\centering
\includegraphics[width=8.0cm]{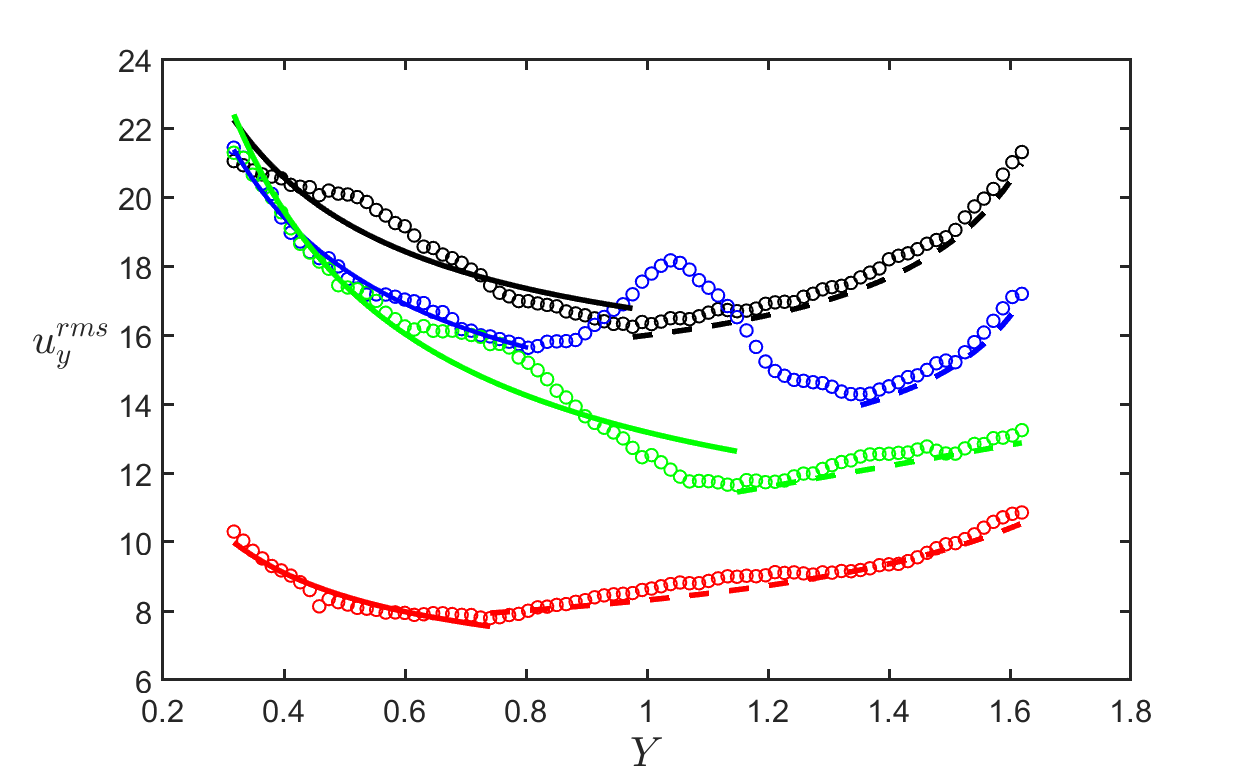}
\includegraphics[width=7.6cm]{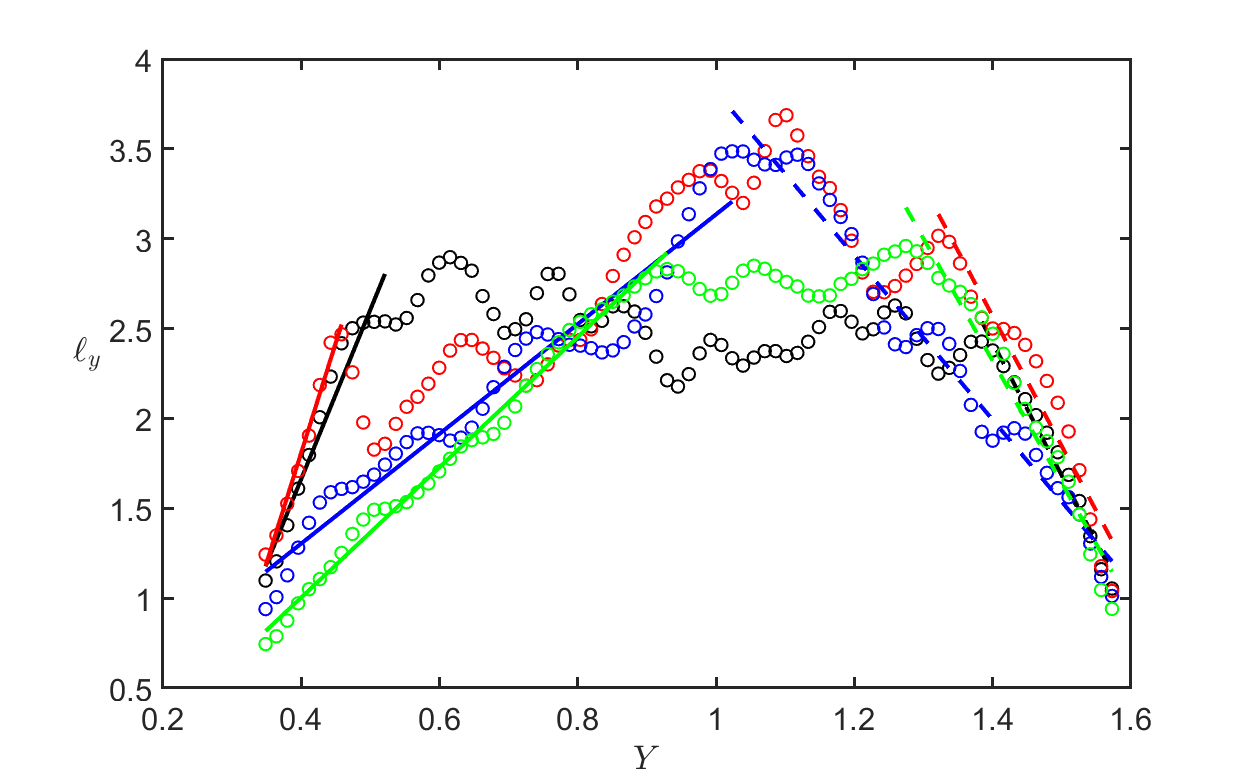}
\caption{\label{Fig7}
Dependencies of  the turbulent velocity $u_y^{\rm rms}(Y)$ (left panel)
and the horizontal  integral turbulent length scale $\ell_y(Y)$ (right panel)
on the horizontal coordinate $Y$ in the core flow averaged over $Z$
in the range $0.4  \leq Z/L_z \leq 0.8$ (in the region of jets) for isothermal turbulence (black)
and for  forced convective turbulence at the temperature differences
$\Delta T = 40$~K (red), $\Delta T = 50$~K  (blue) and $\Delta T = 60$~K  (green)
between the bottom and upper walls of the chamber.
Fitting curves are shown by the solid (left part) and dashed (right part) lines.
The velocity is measured in cm/s and the coordinates $Y$ is normalized by $L_z=26$ cm.
}
\end{figure*}

\begin{figure*}[t!]
\centering
\includegraphics[width=17.0cm]{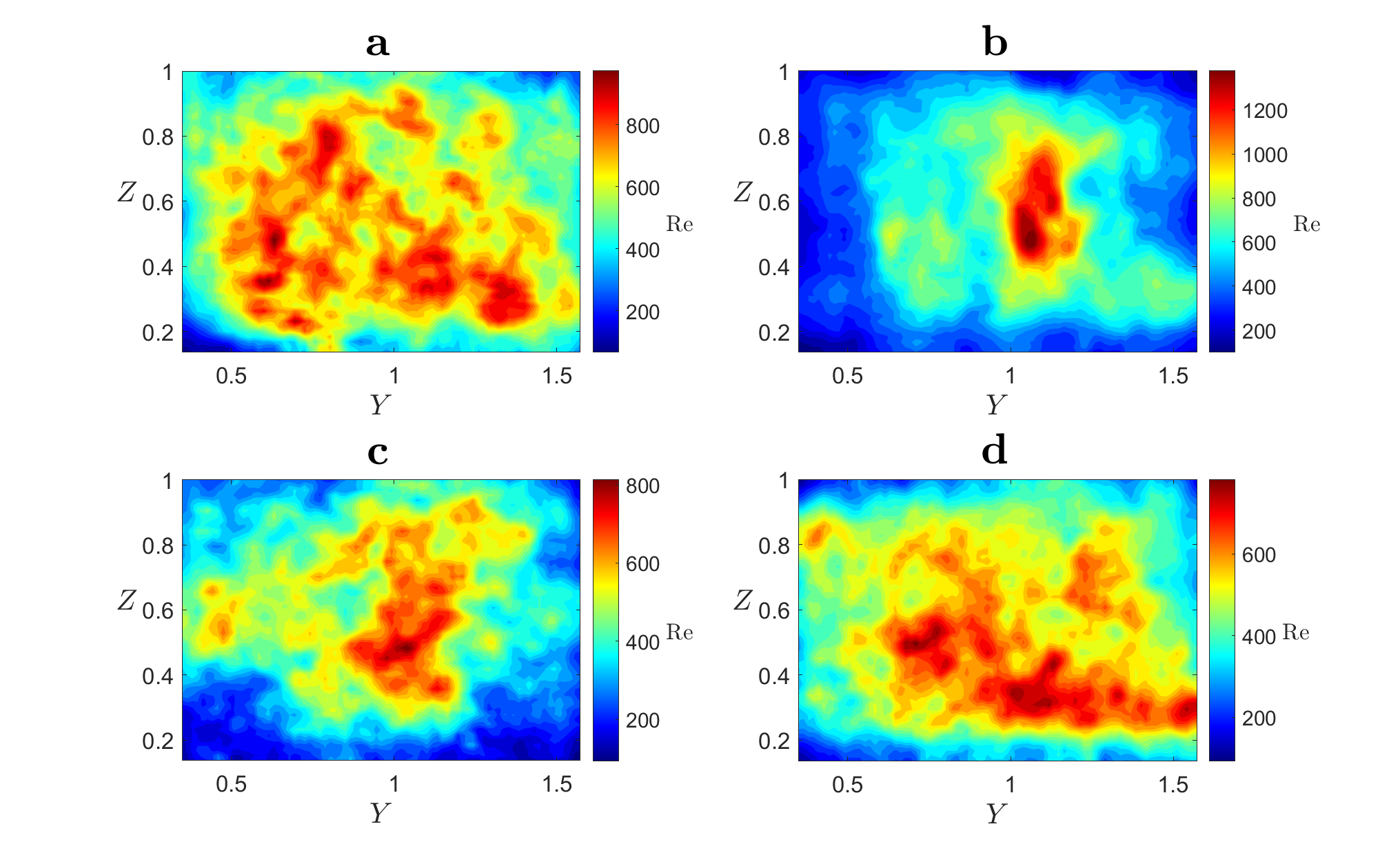}
\caption{\label{Fig8}
Distributions of  the Reynolds number Re$=(u^{\rm (rms)}_y \, \ell_y + 2 u^{\rm (rms)}_z \, \ell_z)/\nu$
in the core flow: {\bf (a)} for isothermal turbulence; and
for forced convective turbulence at the temperature differences:
{\bf (b)} $\Delta T = 40$~K between the bottom and upper walls of the chamber;
{\bf (c)} $\Delta T = 50$~K; and {\bf (d)} $\Delta T = 60$~K.
The coordinates $Y$ and $Z$ are normalized by $L_z=26$ cm.
\\
\\
}
\end{figure*}

\begin{figure*}[t!]
\centering
\includegraphics[width=17.0cm]{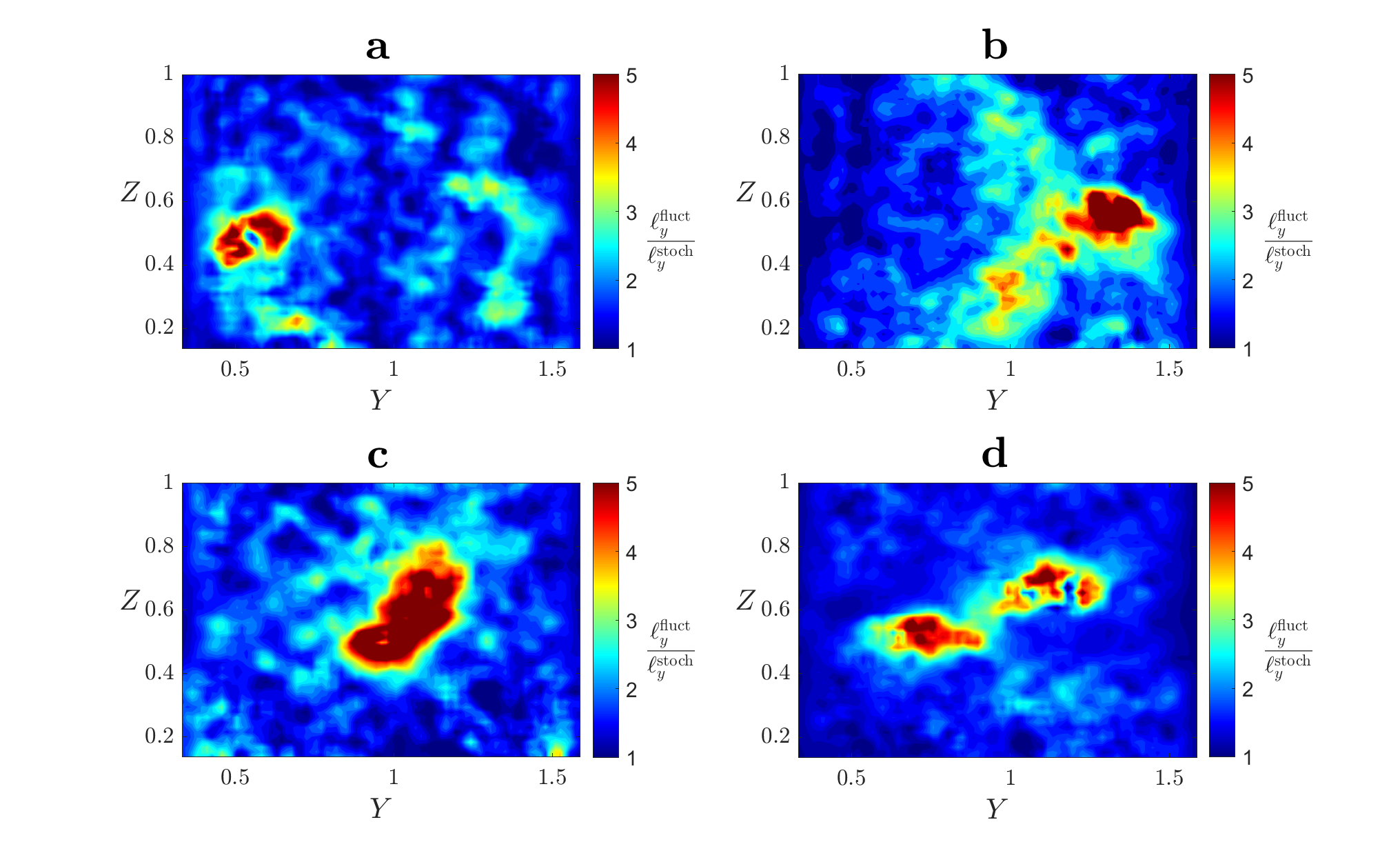}
\caption{\label{Fig10}
Distributions of the ratio of the horizontal  integral turbulent length scale $|\ell_y^{\rm fluct}|$
obtained by the Reynolds averaging to that $|\ell_y^{\rm stoch}|$ of the stochastic part of velocity fluctuations
obtained by the proper orthogonal decomposition
in the core flow: {\bf (a)} for isothermal turbulence; and
for forced convective turbulence at the temperature differences:
{\bf (b)} $\Delta T = 40$~K between the bottom and upper walls of the chamber;
{\bf (c)} $\Delta T = 50$~K; and {\bf (d)} $\Delta T = 60$~K.
The coordinates $Y$ and $Z$ are normalized by $L_z=26$ cm.
}
\end{figure*}

The velocity fields in our experiments are measured  in a flow
domain $358.4 \times 237.5$ mm$^2$ with a spatial
resolution of $2002 \times 1326$ pixels, so that
a spatial resolution 178 $\mu$m /pixel have been achieved.
The velocity field is analysed in the probed region
with interrogation windows of $32\times 32$ pixels.
The velocity measurements in our experiments allow us to determine various  turbulence characteristics:
the mean velocity and the root mean square (r.m.s.)
velocity fluctuations, the turbulent anisotropy,
the two-point correlation functions of velocity fluctuations and integral scales of turbulence
in the horizontal  and vertical directions, as well as
the Reynolds number (see Figs.~\ref{Fig2}--\ref{Fig8}).
We determine the mean and r.m.s. velocities for
every point of a velocity map by averaging over 300 independent maps.
The integral length scales of turbulence $\ell_y$ and $\ell_z$
in the horizontal $Y$ and the vertical $Z$ directions
are determined from the two-point correlation functions of the velocity field.

\begin{figure}[t!]
\centering
\includegraphics[width=8.0cm]{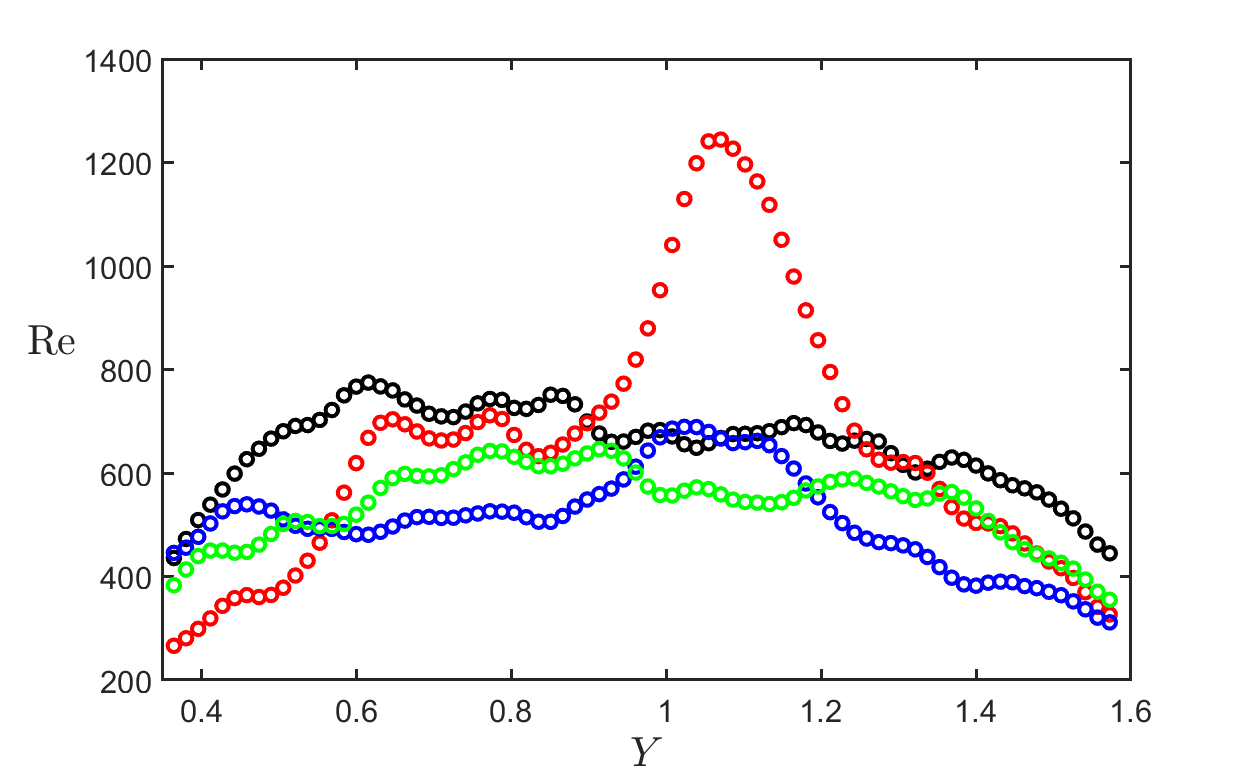}
\caption{\label{Fig9}
Dependence of the Reynolds number ${\rm Re}(Y)$
on the horizontal coordinate $Y$ in the core flow averaged over $Z$
in the range $0.4  \leq Z/L_z \leq 0.8$ (in the region of jets) for isothermal turbulence (black)
and for  forced convective turbulence at the temperature differences
$\Delta T = 40$~K (red), $\Delta T = 50$~K  (blue) and $\Delta T = 60$~K  (green)
between the bottom and upper walls of the chamber.
The coordinates $Y$ is normalized by $L_z=26$ cm.
}
\end{figure}

The particle spatial distribution is obtained by the PIV system
using the effect of the Mie light scattering by particles \cite{guib01}.
In particular, the mean intensity of scattered light is determined
in $43 \times 64$ interrogation windows with the size
$32 \times 32$ pixels.
This allows us to find the vertical and horizontal distributions of the intensity of
the scattered light.
We take into account that the light radiation energy flux scattered
by small particles is given by $E_s \propto E_0 \Psi(\pi d_{\rm p}/\lambda; a_0;n)$.
Here $\Psi$ is the scattering function, $d_{\rm p}$ is the particle diameter, $\lambda$ is the
wavelength, $a_0$ is the index of refraction.
The energy flux incident at the particle is $E_0 \propto \pi d_{\rm p}^2 / 4$.
When $\lambda > \pi d_{\rm p}$, the  scattering function $\Psi$ is obtained
from the Rayleigh's law, $\Psi \propto d_{\rm p}^4$.
For small $\lambda$, the scattering function $\Psi$ is independent of
the particle diameter and the wavelength.
In a general case, the scattering function $\Psi$ is obtained using the Mie
equations \cite{BH83}.

In addition, we take into account that the light radiation energy flux scattered
by small particles is $ E_s \propto E_0 \, n \, (\pi d_{\rm p}^2 / 4) $.
Therefore, the scattered light energy flux
incident on the charge-coupled device (CCD) camera probe is proportional
to the particle number density $n$.
The ratio of the scattered radiation fluxes at two locations in the flow and at the
image measured with the CCD camera is equal to the ratio of the
particle number densities at these two locations.
The scattered light intensity $E^T$ obtained in
a temperature-stratified turbulence is normalized by
the scattered light intensity E measured in the
isothermal case obtained under the same conditions.
Our measurements show that
using different concentrations of the incense smoke,
the distribution of the scattered light intensity averaged over
a vertical coordinate is independent of the particle number
density in the isothermal flow.
Therefore, using this normalization, we can characterize the spatial distribution of particle
number density $n \propto E^T /E_0$ in the temperature stratified turbulence.

The measurement technique and data processing procedure described in this section
are similar to those used by us in various experiments with
turbulent convection \cite{BEKR09,EEKR11,BELR20,SKRL22,EKRL23}
and stably stratified turbulence \cite{EEKR13,CEKR14,EKRL22}.
The similar measurement technique and data processing procedure
in the experiments have been applied previously by us to investigate
the phenomenon of turbulent thermal
diffusion in a homogeneous turbulence \cite{BEE04,EEKR04,EEKR06a,AEKR17},
mixing of particles in inhomogeneous turbulence \cite{EHSR09},
as well as for study of small-scale particle clustering \cite{EKR10}.

\section{Experimental results}
\label{sect3}

In this section we analyse the obtained experimental results
in a convective turbulence forced by two similar turbulence generators
with oscillating membrane and a steady grid  attached to the side walls of the chamber
(see Fig.~\ref{Fig1}).
In Fig.~\ref{Fig2} we show the mean velocity field in the core flow
(excluding near-boundary regions in the chamber)
for  isothermal turbulence (see Fig.~\ref{Fig2}a),
and forced convective turbulence at the temperature differences $\Delta T = 40$~K (see Fig.~\ref{Fig2}b), $\Delta T = 50$~K (see Fig.~\ref{Fig2}c) and $\Delta T = 60$~K (see Fig.~\ref{Fig2}d) between the bottom and upper walls of the chamber.
The mean velocity patterns in these experiments are very complicated containing
one or two large-scale circulations separated by jets produced by
two turbulence generators with oscillating membranes and a steady grids.
In particular, we observe a transition between a single-roll pattern for isothermal
turbulence (Fig.~\ref{Fig2}a) to double-roll pattern with increase of
the temperature differences between the bottom and upper walls of the chamber
(Figs.~\ref{Fig2}b,c),
and at larger temperature differences a single-roll structure  is observed again (Fig.~\ref{Fig2}d).

In these experiments, the turbulent kinetic energy is produced by buoyancy
and the membrane oscillations.
Large-scale shear in such mean velocity fields contributes to the turbulence production rate.
The imposed temperature differences $\Delta T$ between the bottom and upper walls of the chamber
increases buoyancy in convective turbulence which strongly affects the mean velocity patterns.
The turbulence generators in the side walls of the chamber produce jets in opposite directions creating strong shear flows.
Distributions of various turbulent characteristics in the core flow of the chamber
for isothermal and convective turbulence at different temperature differences $\Delta T$
between the bottom and upper walls of the chamber are shown in Figs.~\ref{Fig3}--\ref{Fig8}.
In particular, we plot the distributions of the turbulent velocity $|u^{\rm rms}| = [\langle u_y^2 \rangle + \langle u_z^2 \rangle]^{1/2}$
(Fig.~\ref{Fig3}), the turbulent anisotropy $[\langle u_y^2 \rangle / \langle u_z^2 \rangle]^{1/2}$ (Fig.~\ref{Fig4}),
the horizontal $\ell_y$ and vertical $\ell_z$ integral turbulent length scales (Figs.~\ref{Fig5}--\ref{Fig6}),
and the Reynolds number Re$=(u^{\rm rms}_y \, \ell_y + 2 u^{\rm rms}_z \, \ell_z)/\nu$ (Fig.~\ref{Fig8}).
In Fig.~\ref{Fig7} we also plot the dependencies of  the turbulent velocity $u_y^{\rm rms}(Y)$
and the horizontal  integral turbulent length scale $\ell_y(Y)$
on the horizontal coordinate $Y$ in the core flow averaged over $Z$.

The above expression for the Reynolds number estimate can be derived as follows.
The Reynolds number can be estimated as Re$=\tau_0 \langle {\bm u}^2  \rangle/\nu$,
where $\langle {\bm u}^2  \rangle= \langle u_x^2  \rangle + \langle u_y^2  \rangle+\langle u_z^2  \rangle$
and $\tau_0$ is the correlation turbulent time.
We assume that the correlation turbulent time is the same along $X$, $Y$ and $Z$ directions, i.e., $\tau_0 = \ell_x/u^{\rm rms}_x = \ell_y/u^{\rm rms}_y = \ell_z/u^{\rm rms}_z$ and $u^{\rm rms}_x  \approx u^{\rm rms}_z$.
This yields the estimate for the Reynolds number Re$=(u^{\rm rms}_y \, \ell_y + 2 u^{\rm rms}_z \, \ell_z)/\nu$.

To determine the integral turbulent length scales, we integrate
the two-point correlation function of velocity fluctuations
over the distance $r$ between two points which varies from zero to $r_{\rm min}$, where the
correlation function attains the first minimum \cite{QL03}.
The reason is that the measured two-point correlation function of velocity fluctuations
does not tends to zero for larger $r$ but it oscillates near some non-zero constant due to a strong influence
of large-scale circulations on velocity fluctuations.

For comparison, we also apply other approach to determine the integral turbulent length scales.
In particular, we use the proper orthogonal decomposition \cite{HLBR12,BL93,BBB12},
which allows us to extract from the measured velocity field the mean field,
the part of velocity fluctuations which are strongly affected by the mean field (which characterises the coherent structures)
and stochastic component of velocity fluctuations.
The integral scale of turbulence is determined by a stochastic component of velocity fluctuations.
In Fig.~\ref{Fig10} we show the distributions of the ratio $\ell_y^{\rm fluct} /\ell_y^{\rm stoch}$
of the horizontal  integral turbulent length scale $\ell_y^{\rm fluct}$
obtained by usual Reynolds averaging (where velocity is decomposed into the mean and fluctuations)
to that $\ell_y^{\rm stoch}$ of the stochastic part of velocity fluctuations
obtained by the proper orthogonal decomposition in the core flow for isothermal turbulence and
for forced convective turbulence at the various temperature differences between the bottom and upper walls of the chamber.

As follows from Fig.~\ref{Fig10}, in many regions  both methods yield
the similar spatial distributions of the integral turbulent length scale and
compared values of the ratio $\ell_y^{\rm fluct} /\ell_y^{\rm stoch}$.
Only there are some spatial regions, where this ratio exceeds 1 substantially.
May be this anomaly is related to the fact that in these regions there is a very complicated
mean velocity field and the proper orthogonal decomposition
does not work well there.

Figures~\ref{Fig3}--\ref{Fig9} demonstrate that convective turbulence is strongly inhomogeneous and anisotropic.
The inhomogeneity of the turbulence intensity in the horizontal $Y$ direction (along the jets)
is stronger than in the vertical $Z$ direction.
The isothermal turbulence and convective turbulence are very different, so the
buoyancy and the mean temperature gradients strongly affect the distributions of the turbulent intensity.
In particular, the isothermal turbulence is more anisotropic than the convective turbulence,
 i.e., large-scale circulations decreases anisotropy, especially away from jets.

In Fig.~\ref{Fig7} and Fig.~\ref{Fig9} we show the dependencies of
the turbulent velocity $u_y^{\rm rms}(Y)$ (the left panel of Fig.~\ref{Fig7}),
the horizontal  integral turbulent length scale $\ell_y(Y)$ (the right panel of Fig.~\ref{Fig7}) and
the Reynolds number ${\rm Re}(Y)$ (Fig.~\ref{Fig9})
on the horizontal coordinate $Y$ in the core flow averaged over vertical coordinate $Z$
in the range $0.4  \leq Z/L_z \leq 0.8$ (in the region of jets created by the two turbulence generators)
for isothermal turbulence (black) and for  forced convective turbulence at the temperature differences
$\Delta T = 40$~K (red), $\Delta T = 50$~K  (blue) and $\Delta T = 60$~K  (green)
between the bottom and upper walls of the chamber.

In various cases of isothermal and forced convection, we observe
the scalings $u_y^{\rm rms}(Y) \propto Y^{-1}$ and $\ell_y(Y) \propto Y$
in the left and right ranges of the horizontal coordinate $Y$ (see Fig.~\ref{Fig7}).
For instance, the data in Fig.~\ref{Fig7} (right panel) are fitted to linear function (solid line is for left region up to first maximum
and dashed line for left from the last maximum). This yields the following profiles for $\ell_y(Y)$:
(a) for isothermal turbulence $\ell_y (Y)=9.4Y-2.1$ (left region) and $\ell_y (Y)=-7.4Y+12.8$ (right region);
(b) for	$\Delta T = 40$~K , $\ell_y (Y)=12.2Y-3.1$ (left region) and $\ell_y (Y)=-7.2Y+12.7$ (right region);
(c) for $\Delta T = 50$~K, $\ell_y (Y)=3.1Y+0.1$ (left region) and $\ell_y (Y)=-4.6Y+8.4$ (right region);
(d) for $\Delta T = 60$~K, $\ell_y (Y)=3.6Y-0.4$ (left region) and $\ell_y (Y)=-6.8Y+11.4$ (right region).
Similarly, the following fitted profiles are obtained for the velocity $u_y^{\rm rms}(Y)$:
(a) for isothermal turbulence $u_y^{\rm rms}(Y)=2.6 Y^{-1} + 14.1$ (left region)
and $u_y^{\rm rms}(Y)=1.7 (1.9-Y)^{-1} + 14.1$ (right region);
(b) for	$\Delta T = 40$~K, $u_y^{\rm rms}(Y)=1.4 Y^{-1} + 5.7$ (left region)
and $u_y^{\rm rms}(Y)=2.7 (2.2-Y)^{-1} + 6.3$ (right region);
(c) for $\Delta T = 50$~K, $u_y^{\rm rms}(Y)=3 Y^{-1} + 12$ (left region)
and $u_y^{\rm rms}(Y)=1.1 (1.8-Y)^{-1} + 12$ (right region);
(d) for $\Delta T = 60$~K, $u_y^{\rm rms}(Y)=4.3 Y^{-1} + 9$ (left region)
and $u_y^{\rm rms}(Y)=2.2 (2.7-Y)^{-1} + 8.1$ (right region).

These scalings resemble qualitatively the results
of the early laboratory experiments conducted with one oscillating grid in isothermal turbulence  \cite{turn68,turn73,tho75,hop76,kit97,san98,med01}
and forced convective turbulence \citep{EKRL22,EKRL23},
where the r.m.s. velocity behaves as $\sqrt{\langle {\bf u'}^2 \rangle} \propto f \, Y^{-1}$,
while the horizontal integral turbulence length scale increases linearly with the distance $Y$ from a grid.

We also study the phenomenon of turbulent thermal diffusion
in a forced convective turbulence by measuring the spatial distributions
of the mean temperature and the mean particle number density.
The initial spatial distributions of small solid particles injected into the chamber,
are nearly homogeneous and isotropic.
In a temperature-stratified turbulence, the spatial distributions of the mean particle number density
is expected to be strongly inhomogeneous due to the effective pumping velocity ${\bm V}^{\rm eff}$
that is proportional $ \propto - D_T \, {\bm \nabla} \ln \overline{T}$, so particles should be accumulated
at the vicinity of the mean temperature minimum.

\begin{figure*}[t!]
\centering
\includegraphics[width=8.0cm]{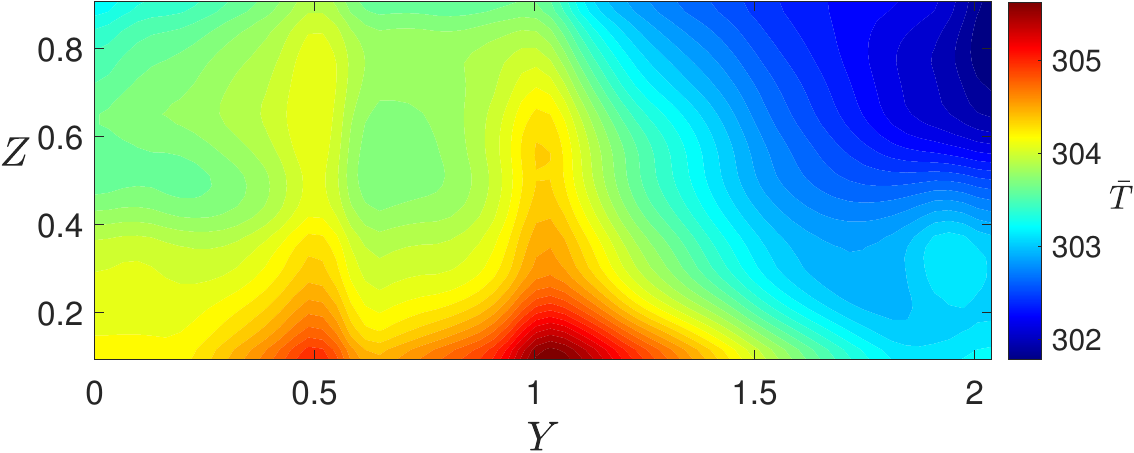}
\includegraphics[width=8.0cm]{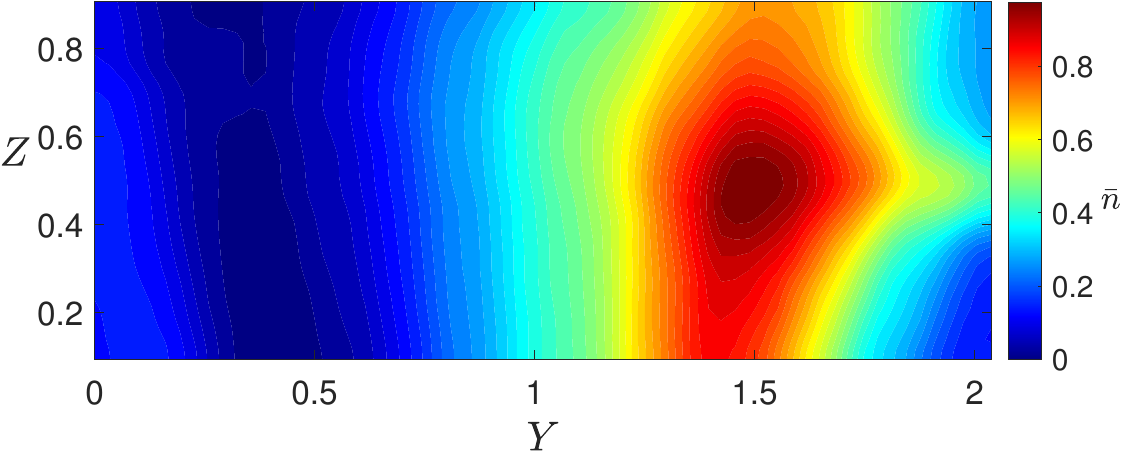}
\caption{\label{Fig11}
Distributions of  the mean temperature $\meanT(Y,Z)$ (left panel)
and the normalized mean particle number density $\meanN(Y,Z) / \meanN_0$ (right panel)
in the core flow for  forced convective turbulence at the temperature difference $\Delta T = 40$~K
between the bottom and upper walls of the chamber.
The coordinates $Y$ and $Z$ are normalized by $L_z=26$ cm.
\\
\\
}
\end{figure*}

\begin{figure*}[t!]
\centering
\includegraphics[width=8.0cm]{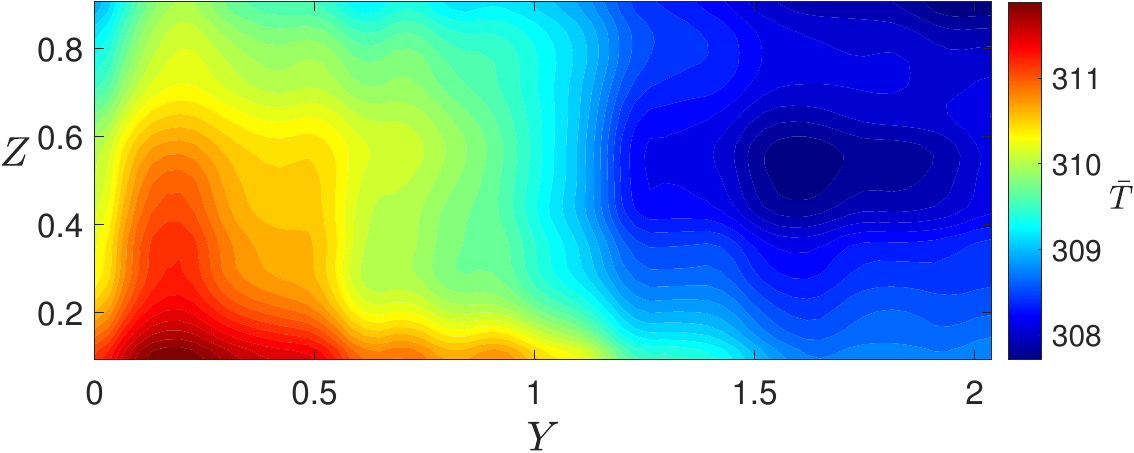}
\includegraphics[width=8.0cm]{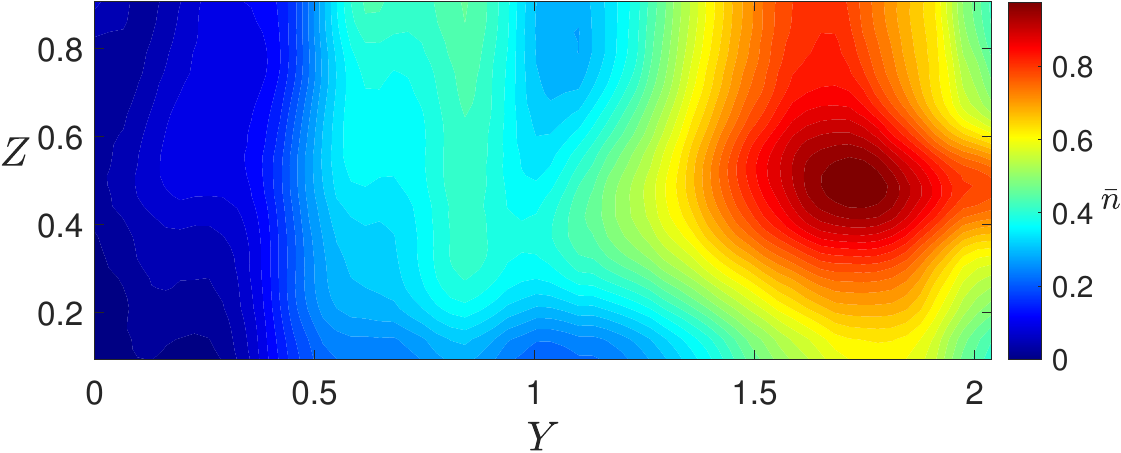}
\caption{\label{Fig12}
Distributions of  the mean temperature $\meanT(Y,Z)$ (left panel)
and the normalized mean particle number density $\meanN(Y,Z) / \meanN_0$ (right panel)
in the core flow for  forced convective turbulence at the temperature difference
$\Delta T = 50$~K between the bottom and upper walls of the chamber.
The coordinates $Y$ and $Z$ are normalized by $L_z=26$ cm.
\\
\\
}
\end{figure*}

\begin{figure*}[t!]
\centering
\includegraphics[width=8.0cm]{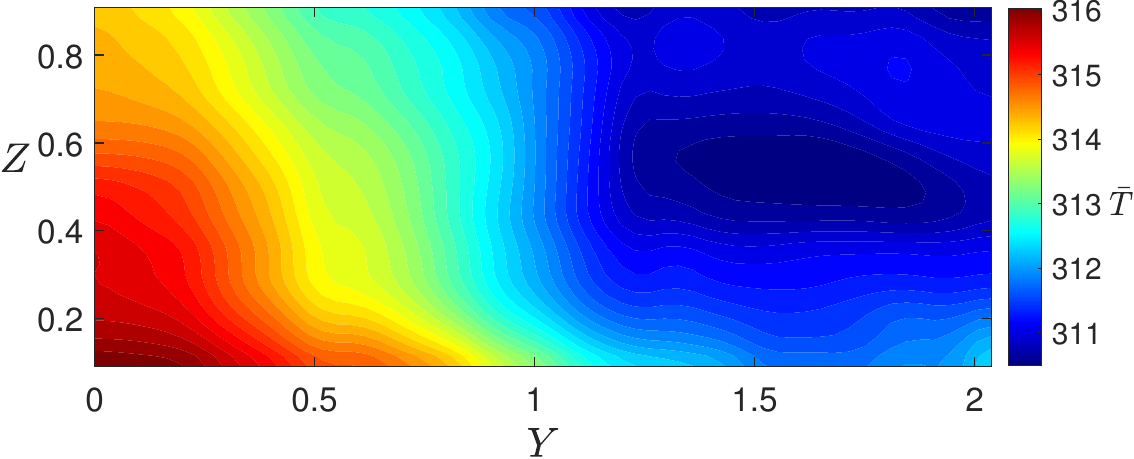}
\includegraphics[width=8.0cm]{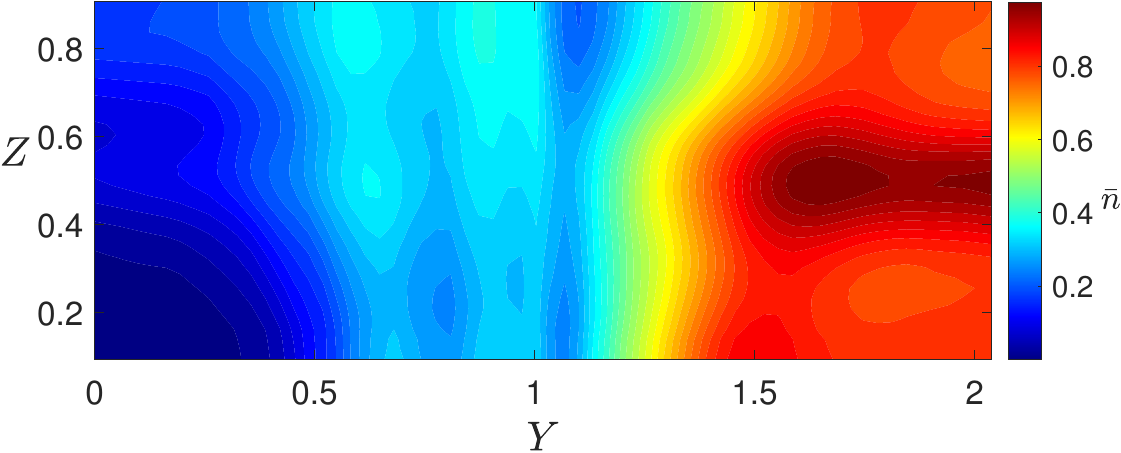}
\caption{\label{Fig13}
Distributions of  the mean temperature $\meanT(Y,Z)$ (left  panel)
and the normalized mean particle number density $\meanN(Y,Z)$ (right panel)
in the core flow for  forced convective turbulence at the temperature difference
$\Delta T = 60$~K between the bottom and upper walls of the chamber.
The coordinates $Y$ and $Z$ are normalized by $L_z=26$ cm, and the mean particle number density $\meanN(Y,Z)$
is normalised by the maximum mean particle number density $\meanN_0$.
}
\end{figure*}

In left panels of Figs.~\ref{Fig11}--\ref{Fig13}, we show the distributions of  the mean temperature $\meanT(Y,Z)$
in the core flow for forced convective turbulence at the temperature difference $\Delta T = 40$~K (see Fig.~\ref{Fig11}),
$\Delta T = 50$~K (see Fig.~\ref{Fig12})
and $\Delta T = 60$~K (see Fig.~\ref{Fig13}) between the bottom and upper walls of the chamber.
We see that the gradients of the mean temperature in the horizontal $Y$ direction (along the jets)
are much more stronger than in the vertical $Z$ direction.

\begin{figure*}[t!]
\centering
\includegraphics[width=18.5cm]{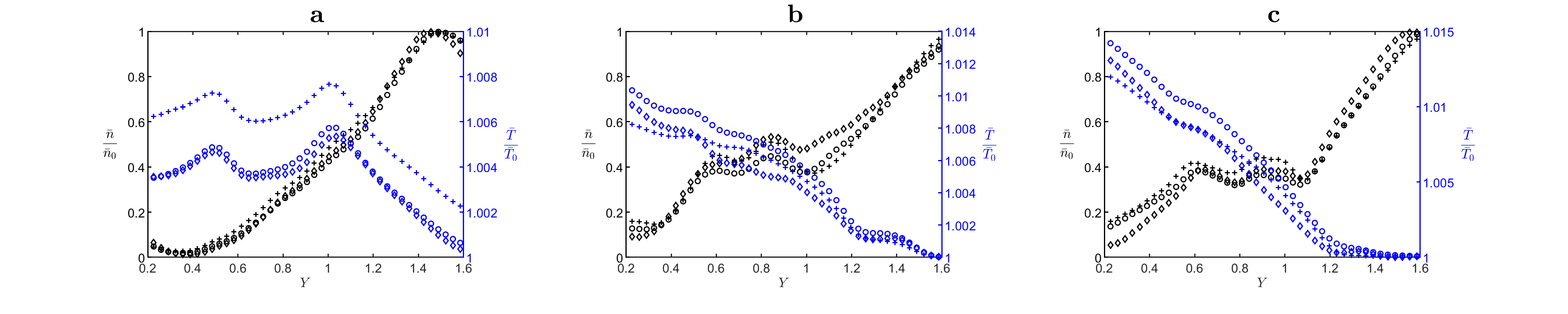}
\caption{\label{Fig14}
Dependencies of  the mean temperature $\meanT(Y)/\meanT_0$
and the normalized mean particle number density $\meanN(Y)/\meanN_0$
on the horizontal coordinate $Y$ in the core flow
for three fixed vertical positions: $Z/L_z=0.23$ (diamonds), $Z/L_z=0.4$ (circles), $Z/L_z=0.58$ (crosses),
for  forced convective turbulence at the temperature differences: {\bf (a)}
$\Delta T = 40$~K; {\bf (b)} $\Delta T = 50$~K and {\bf (c)} $\Delta T = 60$~K
between the bottom and upper walls of the chamber.
The coordinates $Y$ is normalized by $L_z=26$ cm.
}
\end{figure*}

In right panels of Figs.~\ref{Fig11}--\ref{Fig13}, we show the distributions of
the normalized mean particle number density $\meanN(Y,Z) / \meanN_0$.
As expected, the gradients of the particle number density in the horizontal direction
are much more stronger than in the vertical direction.
In the regions with the strong mean temperature gradients,
the maximum of the mean particle number density is located in the regions with minimum
of the mean temperature $\meanT(Y,Z)$ in accordance with the
phenomenon of turbulent thermal diffusion (see Figs.~\ref{Fig11}--\ref{Fig13}).

In Fig.~\ref{Fig14} we plot the dependencies of  the mean temperature $\meanT(Y)/\meanT_0$
and the normalized mean particle number density $\meanN(Y)/\meanN_0$
on the horizontal coordinate $Y$ in the core flow for three fixed vertical positions: $Z/L_z=0.23$ (diamonds), $Z/L_z=0.4$ (circles), $Z/L_z=0.58$ (crosses), for  forced convective turbulence at different temperature differences
between the bottom and upper walls of the chamber.
In spite of the presence of the large-scale circulations superimposed to the convective turbulence,
Fig.~\ref{Fig14} demonstrates the tendency of the localization of the maximum
of the mean particle number density in the vicinity of the minimum in the mean temperature
distribution, and vise versa.
However, this conclusion is only valid in the turbulent regions with large mean temperature gradients.
For these regions, the dominant effect of large-scale particle clustering is turbulent thermal diffusion.
Deviations from this trend occurs in the regions with strong mean fluid motions
where the mean temperature gradient is small.

Note that the terminal fall velocity for the micron-size particles is about $10^{-2}$ cm/s,
while the turbulent velocity in the experiments
with the forced convective turbulence is much larger than
the particle terminal fall velocity
(it is about $25$ cm/s near the turbulence generator and is more than $10$ cm/s far from
the turbulence generator).
This implies that sedimentation of particles does not play any role
in a formation of inhomogeneous particle distributions.
The effective pumping velocity caused by turbulent
thermal diffusion is about $1$--$5$ cm/s.
This implies that the turbulent and effective pumping velocities
in the experiments are much larger than the terminal fall velocity
for micron-size particles.

\section{Conclusions}
\label{sect4}

Properties of turbulence and turbulent transport of non-inertial particles have been studied
in experiments in a turbulent convection forced by two similar turbulence generators at the side walls of the chamber with oscillating
membrane and a steady grid in the air flow with the Rayleigh number about $10^8$.
Particle Image Velocimetry system has been used for measurements of velocity fields and mean particle number density
spatial distributions, while the temperature field has been measured in many locations
applying a temperature probe equipped with 12 E - thermocouples.

The observed large-scale mean flow patterns are formed by an interaction
between the Rayleigh–B\'{e}nard convective cells and jets produced by the turbulence generators.
The mean fluid flow patterns show transition between a single-roll structure for isothermal
turbulence to double-roll mean flow structures with increase of
the temperature differences $\Delta T$ between the bottom and upper walls of the chamber,
and at larger temperature differences a single-roll structure  is observed again.

This picture is opposite to that observed in a free convection \cite{EEKR06c},
where two-roll patterns for small temperature differences $\Delta T$
is replaced by a single-roll pattern with increasing $\Delta T$.
Moreover, the two-roll patterns is recovered for large temperature differences $\Delta T$
in a free convection.
Turbulence in the forced convection is produced by buoyancy and large-scale velocity
shear of large-scale circulations and colliding jets from the side walls of the system.

In the turbulent regions with large mean temperature gradient,
the dominant effect of large-scale particle clustering is turbulent thermal diffusion
so that the distributions of the mean particle number density
are anti-correlated with the mean temperature distributions.
This implies that the maximum particle number density is located in the regions
with the minimum mean fluid temperature, which is qualitatively consistent with
the predictions based on the phenomenon of turbulent thermal diffusion.
However, deviations from this trend occurs in the regions with strong mean fluid motions
where the mean temperature gradient is small.

Note that we have performed experiments with small non-inertial particles.
For inertial particles a large-scale clustering can be more stronger,
because there is an additional positive contribution to the effective pumping velocity
(see, e.g., Ref.~\cite{RI21}).
However, it is very difficult to perform experiments with inertial particles
that is a subject of a separate investigation.

In the present study we have done only a qualitative comparison between theoretical predictions of large-scale particle clustering and experimental results. Since we measure a two-dimensional velocity field (in the $YZ$ plane), and since the large-scale complicated three-dimensional flow contributes to the large-scale clustering of particles in the turbulent
regions with small mean temperature gradients, it is not visible to perform a quantitative
comparisons between theoretical and experimental results.

\bigskip
\noindent
{\bf ACKNOWLEDGEMENTS}
\medskip

We are thankful to the two referees for providing constructive comments
that improved our paper.

\bigskip
\noindent
{\bf AUTHOR DECLARATIONS}

\medskip
{\bf  Conflict of Interest}

The authors have no conflicts to disclose.

\bigskip
\noindent
{\bf DATA AVAILABILITY}
\medskip

The data that support the findings of this study are available from the corresponding author
upon reasonable request.

 \appendix

 \section{Derivation of the turbulent flux of  particles}

 To find the particle turbulent flux, we use an equation for particle number density fluctuations $n'$, that is obtained by subtracting Eq.~(\ref{W2}) from Eq.~(\ref{W1}), which yields
\begin{eqnarray}
{\partial n' \over \partial t} + {\bm \nabla}{\bf \cdot} \left(n' \, {\bm u} - \langle n' \, {\bm u} \rangle \right) - D  \Delta n' = -({\bm u} {\bf \cdot} {\bm \nabla}) \overline{n} -\overline{n} ({\bm \nabla} {\bf \cdot} {\bm u}) ,
\nonumber\\
\label{W3}
\end{eqnarray}
where ${\cal Q} = {\bm \nabla}{\bf \cdot} \left(n' \, {\bm u} - \langle n' \, {\bm u} \rangle \right)$ is the nonlinear term, and
$-({\bm u} {\bf \cdot} {\bm \nabla}) \overline{n} -\overline{n} ({\bm \nabla} {\bf \cdot} {\bm u})$ are the source terms for particle number density fluctuations.
Here, for simplicity, we assumed that mean velocity $\meanUU$ vanishes.
The fluid density is decomposed into the mean fluid density $\overline{\rho}$ and fluctuations $\rho'$, and for a low-Mach-number flow, $|\rho'| \ll \overline{\rho}$.
The anelastic approximation yields
$\bec{\nabla} {\bf \cdot} {\bm u} \approx - \overline{\rho}^{\, -1} \, ({\bm u} \cdot {\bm \nabla}) \,\overline{\rho} ={\bm u} \cdot {\bm \lambda}$,
where ${\bm \lambda} = - {\bm \nabla} \overline{\rho} / \overline{\rho}$.
Therefore, $-\overline{n} ({\bm \nabla} {\bf \cdot} {\bm u}) = (\overline{n}/\overline{\rho} \,) ({\bm u} {\bf \cdot} {\bm \nabla}) \overline{\rho} = - \overline{n} \, ({\bm u} {\bf \cdot} {\bm \lambda})$.
The ratio of the absolute values of the nonlinear term $|{\cal Q}|$ to the diffusion term $|D \Delta n'|$ is the P\'{e}clet number for particles,
that can be estimated as ${\rm Pe} = u_0 \, \ell_0 / D$.

Note that nonlinear equation~(\ref{W3}) cannot be solved exactly for arbitrary P\'{e}clet numbers.
We consider a one-way coupling, and take into account the effect of the turbulent velocity on the particle number density.
On the other hand, we neglect the feedback effect of the particle number density on the turbulent fluid flow.
This one-way coupling approximation is valid when the spatial density of particles $n \, m_p$ is much smaller than the fluid density $\rho$, where $m_p$ is the particle mass.

We apply for simplicity the dimensional analysis to solve Eq.~(\ref{W3}). The dimension of
the left-hand side of Eq.~(\ref{W3}) is the rate of change of particle number density fluctuations $n' / \tau_{n'}$, where $\tau_{n'}$ is the characteristic time of particle number density fluctuations.
For large Reynolds and P\'{e}clet numbers, the characteristic time of particle number density fluctuations $\tau_{n'}$
can be identified with the correlation time $\tau_0$ of the turbulent velocity field.
This implies that in the framework of the dimensional analysis, we replace the left-hand side of Eq.~(\ref{W3}) by $n' / \tau_0$, that yields:
\begin{eqnarray}
n' = - \tau_0 \, \left[({\bm u} {\bf \cdot} {\bm \nabla}) \overline{n} + \overline{n} \, ({\bm \nabla} {\bf \cdot} {\bm u})\right].
\label{W4}
\end{eqnarray}
Multiplying Eq.~(\ref{W4}) by velocity fluctuations, $u_i$, and averaging
over an ensemble of turbulent velocity field, we obtain the turbulent flux of particles as
\begin{eqnarray}
\left\langle n' \, u_i \right\rangle &=& - \tau_0 \,\left\langle u_i u_j  \right\rangle \, \nabla_j \overline{n} - \tau_0 \, \overline{n} \,\left\langle u_i (\bec{\nabla} {\bf \cdot} {\bm u}) \right\rangle
\nonumber\\
&\equiv& V_i^{\rm eff} \, \overline{n} - D_{ij}^{(n)} \, \nabla_j \overline{n} .
\label{W5}
\end{eqnarray}
The last term, $- D_{ij}^{(n)} \, \nabla_j \overline{n} $, in
Eq.~(\ref{W5}) describes the contribution to the flux of particles caused by
turbulent diffusion, where
$D_{ij}^{(n)} = \tau_0 \,\left\langle u_i u_j  \right\rangle$
is the turbulent diffusion tensor.
For an isotropic turbulence
$\langle u_i u_j  \rangle = \delta_{ij} \, \langle {\bm u}^2  \rangle/3$,
so that the turbulent diffusion tensor for large P\'{e}clet numbers is given by
$D_{ij}^{(n)} =D_{\rm T} \delta_{ij}$, where $D_{\rm T}= \tau_0 \,\left\langle {\bm u}^2  \right\rangle / 3$
is the turbulent diffusion coefficient and $\delta_{ij}$ is the unit Kronecker tensor
(i.e., the diagonal components of the tensor $\delta_{ii}=1$, and the off-diagonal components are zero).

The term ${\bm V}^{\rm eff} \, \overline{n}$ in
Eq.~(\ref{W5}) determines the contribution to the turbulent flux of particles caused by
the effective pumping velocity:
${\bm V}^{\rm eff} = - \tau_0 \,\left\langle {\bm u} (\bec{\nabla} {\bf \cdot} {\bm u}) \right\rangle$.
Applying the anelastic approximation,
${\bm \nabla} {\bf \cdot} {\bm u} = - (1/\overline{\rho}) \, ({\bm u} {\bf \cdot} {\bm \nabla}) \overline{\rho}$, we obtain
the effective pumping velocity as
$V_i^{\rm eff} = - \tau_0 \, \left\langle u_i u_j \right \rangle \, \lambda_j  $, where
${\bm \lambda} = - {\bm \nabla} \overline{\rho} / \overline{\rho}$.
For an isotropic turbulence, the effective pumping velocity is given by \cite{EKR96,EKR97}
\begin{eqnarray}
{\bm V}^{\rm eff}  = D_{\rm T} \, {{\bm \nabla} \overline{\rho} \over \overline{\rho}},
\label{NW7}
\end{eqnarray}
and the particle turbulent flux $\langle  {\bm u} \, n'  \rangle$ is
\begin{eqnarray}
\left\langle {\bm u} \, n'  \right\rangle = {\bm V}^{\rm eff} \, \overline{n} - D_{\rm T} \, {\bm \nabla} \overline{n} .
\label{NW8}
\end{eqnarray}

Let us disscuss the physics related to the effective pumping velocity ${\bm V}^{\rm eff}$.
To this end, we use the equation of state for a perfect gas,
$P=(k_B/ m_\mu) \, \rho \, T$, that can be also rewritten for the mean fields as
$\overline{P}=(k_B/ m_\mu) \, \overline{\rho} \, \overline{T}$.
Here $k_B=R/N_A$ is the Boltzmann constant, $R$ is the gas constant,
$N_A$ is the Avogadro number, $\mu=m_\mu N_A$ is the molar mass,
and $\overline{P}$ and $\meanT$ are the mean pressure
and mean temperature, respectively.
We assume that $\overline{\rho} \, \meanT \gg \langle \rho' \, \theta \rangle$,
and express the gradient of the mean fluid density in terms of the gradients of
the mean fluid pressure ${\bm \nabla} \overline{P}$ and mean fluid temperature ${\bm \nabla} \meanT$ as
${\bm \nabla} \, \ln\overline{\rho} = {\bm \nabla} \ln \overline{P} - {\bm \nabla} \ln \meanT$.
For small mean pressure gradient, ${\bm \nabla} \, \ln\overline{\rho} \approx  - {\bm \nabla} \ln \meanT$,
and the effective pumping velocity for non-inertial particles is given by \cite{EKR96,EKR97}
\begin{eqnarray}
{\bm V}^{\rm eff} = - D_{\rm T} \, {{\bm \nabla} \meanT \over \meanT} .
\label{NW9}
\end{eqnarray}


\end{document}